\documentclass[fleqn]{2017SCGE}
\usepackage{float}
\begin{document}
\ensubject{subject}
\ArticleType{Article}
\SpecialTopic{SPECIAL TOPIC: }
\Year{2021}
\Month{March}
\Vol{00}
\No{0}
\DOI{10.1007/00000}
\ArtNo{000000}
\ReceiveDate{xxx xx, 2021}
\AcceptDate{xxx xx, 2021}

\title{Constraining self-interacting dark matter with the full dataset of PandaX-II}

\address[ ]{(PandaX-II Collaboration)}
\author[1]{Jijun Yang}{}
\author[1]{Abdusalam Abdukerim}{}
\author[1]{Wei Chen}{}
\author[1,2]{Xun Chen}{}
\author[3]{Yunhua Chen}{}
\author[4]{Chen Cheng}{}
\author[5]{Xiangyi Cui}{}
\author[6]{\\Yingjie Fan}{}
\author[7]{Deqing Fang}{}
\author[7]{Changbo Fu}{}
\author[8]{Mengting Fu}{}
\author[9,10]{Lisheng Geng}{}
\author[1]{Karl Giboni}{}
\author[1]{Linhui Gu}{}
\author[3]{\\Xuyuan Guo}{}
\author[1]{Ke Han}{}
\author[1]{Changda He}{}
\author[3]{Shengming He}{}
\author[1]{Di Huang}{}
\author[3]{Yan Huang}{}
\author[11]{Ran Huo}{}
\author[12]{Yanlin Huang}{}
\author[1]{\\Zhou Huang}{}
\author[13]{Xiangdong Ji}{}
\author[14]{Yonglin Ju}{}
\author[5]{Shuaijie Li}{}
\author[15,16]{Qing Lin}{}
\author[14]{Huaxuan Liu}{}
\author[1,2,5]{Jianglai Liu\footnote{Spokesperson: jianglai.liu@sjtu.edu.cn}}{}
\author[17,18]{\\Xiaoying Lu}{}
\author[1]{Wenbo Ma}{}
\author[7]{Yugang Ma}{}
\author[8]{Yajun Mao}{}
\author[1,2]{Yue Meng}{}
\author[17,18]{Nasir Shaheed}{}
\author[1]{Kaixiang Ni}{}
\author[3]{\\Jinhua Ning}{}
\author[1]{Xuyang Ning}{}
\author[17,18]{Xiangxiang Ren}{}
\author[3]{Changsong Shang}{}
\author[9]{Guofang Shen}{}
\author[1]{Lin Si}{}
\author[13]{\\Andi Tan}{}
\author[17,18]{Anqing Wang}{}
\author[19]{Hongwei Wang}{}
\author[17,18]{Meng Wang}{}
\author[7]{QiuHong Wang}{}
\author[8]{Siguang Wang}{}
\author[4]{\\Wei Wang}{}
\author[14]{Xiuli Wang}{}
\author[1,2]{Zhou Wang}{}
\author[4]{Mengmeng Wu}{}
\author[3]{Shiyong Wu}{}
\author[1]{Weihao Wu}{}
\author[1]{\\Jingkai Xia}{}
\author[13,21]{Mengjiao Xiao}{}
\author[4]{Xiang Xiao}{}
\author[5]{Pengwei Xie}{}
\author[1]{Binbin Yan}{}
\author[1]{Yong Yang\footnote{Corresponding author: yong.yang@sjtu.edu.cn}}{}
\author[6]{\\Chunxu Yu}{}
\author[20]{Hai-Bo Yu\footnote{Corresponding author: haiboyu@ucr.edu}}{}
\author[17,18]{Jumin Yuan}{}
\author[1]{Ying Yuan}{}
\author[1]{Xinning Zeng}{}
\author[13]{Dan Zhang}{}
\author[1]{Tao Zhang}{}
\author[1]{\\Li Zhao}{}
\author[10]{Qibin Zheng}{}
\author[3]{Jifang Zhou}{}
\author[1]{Ning Zhou}{}
\author[9]{Xiaopeng Zhou}{}

\AuthorMark{Yang J J}
\AuthorCitation{Yang J J, et al}

\address[1]{ School of Physics and Astronomy, Shanghai Jiao Tong University, MOE Key Laboratory for Particle \\
  Astrophysics and Cosmology, Shanghai Key Laboratory for article Physics and Cosmology, Shanghai 200240, China}
\address[2]{ Shanghai Jiao Tong University Sichuan Research Institude, Chengdu 610213, China}
\address[3]{ Yalong River Hydropower Development Company, Ltd., 288 Shuanglin Road, Chengdu 610051, China}
\address[4]{ School of Physics, Sun Yat-Sen University, Gaungzhou 510275, China}
\address[5]{ Tsung-Dao Lee Institute, Shanghai 200240, China}
\address[6]{ School of Physics, Nankai University, Tianjin 300071, China}
\address[7]{ Key Laboratory of Nuclear Physics and Ion-beam Application (MOE),\\
  Institute of Modern Physics, Fudan University, Shanghai 200433, China}
\address[8]{ School of Physics, Peking University, Beijing 100871, China}
\address[9]{ School of Physics, Beihang University, Beijing 100191, China}
\address[10]{ International Research Center for Nuclei and Particles in the Cosmos \& Beijing Key Laboratory\\
  of Advanced Nuclear Materials and Physics, Beihang University, Beijing 100191, China}
\address[11]{ Shandong Institute of Advanced Technology, Jinan 250103, China}
\address[12]{ School of Medical Instrument and Food Engineering,\\
  University of Shanghai for Science and Technology, Shanghai 200093, China}
\address[13]{ Department of Physics, University of Maryland, College Park, Maryland 20742, USA} 
\address[14]{ School of Mechanical Engineering, Shanghai Jiao Tong University, Shanghai 200240, China}
\address[15]{ State Key Laboratory of Particle Detection and Electronics,\\
  University of Science and Technology of China, Hefei 230026, China}
\address[16]{ Department of Modern Physics, University of Science and Technology of China, Hefei 230026, China}
\address[17]{ Key Laboratory of Particle Physics and Particle Irradiation of Ministry\\
  of Education, Shandong University, Jinan 250100, China}
\address[18]{ Research Center for Particle Science and Technology, Institute of Frontier\\
  and Interdisciplinary Scienc, Shandong University, Qingdao 266237, Shandong, China }
\address[19]{ Shanghai Advanced Research Institute, Chinese Academy of Sciences, Shanghai 201210, China}

\address[20]{ Department of Physics and Astronomy, University of California, Riverside, California 92507, USA}
\address[21]{ Center of High Energy Physics, Peking University, Beijing 100871, China}

\abstract{Self-interacting Dark Matter (SIDM) is a leading candidate
  proposed to solve discrepancies between predictions of the
  prevailing cold dark matter theory and observations of
  galaxies. Many SIDM models predict the existence of a light force
  carrier that mediate strong dark matter self-interactions. If the
  mediator couples to the standard model particles, it could produce
  characteristic signals in dark matter direct detection
  experiments. We report searches for SIDM models with a light
  mediator using the full dataset of the PandaX-II experiment, based
  on a total exposure of 132 tonne-days. No significant excess over
  background is found, and our likelihood analysis leads to a strong
  upper limit on the dark matter-nucleon coupling strength. We further
  combine the PandaX-II constraints and those from observations of the
  light element abundances in the early universe, and show that direct
  detection and cosmological probes can provide complementary constraints on dark matter models with a light mediator.}

  
\maketitle
\begin{multicols}{2}

  \section{Introduction}
  \label{sec:1}

A fundamental question beyond the standard model of particle physics
concerns the nature of dark matter, which makes up over $80\%$ of the
mass in the universe. In the prevailing theory, dark matter consists
of cold and collisionless particles, for example, in the form of
weakly interacting massive particles
(WIMPs)~\cite{BERTONE2005279}. This theory is extremely successful in
explaining large-scale structure of the
universe~\cite{Aghanim:2018eyx}. But it has difficulties in
accommodating observations of dark matter distributions in galaxies,
see~\cite{Tulin:2017ara,Bullock:2017xww}. A promising solution is to
consider self-interacting dark matter
(SIDM)~\cite{Spergel:1999mh,Kaplinghat:2015aga}, where dark matter
particles have strong self-interactions. Studies show that SIDM can
change the inner halo structure and provide better agreement with
galactic observations, see~\cite{Tulin:2017ara} for a review and
references therein.
  
In many particle physics realizations of SIDM, a light force carrier
($\phi$) is introduced to mediate the self-interactions and its
typical mass is $\sim10~{\rm MeV}$~\cite{Tulin:2017ara}. When $\phi$
couples to the standard model particles, it can lead to characteristic
signals in direct detection experiments such as
PandaX-II~\cite{Kaplinghat:2013yxa}. Compared to traditional WIMP
searches, the SIDM signal spectrum is peaked more towards low energies
since the mediator mass is comparable to the momentum transfer in
nuclear recoils~\cite{DelNobile:2015uua,Li:2014vza}. For the same reason, direct
searches can put a strong constraint on the mixing parameter between
the two sectors. In~\cite{Ren:2018gyx}, we reported
searches for dark matter-nucleon interactions mediated by a light
mediator using PandaX-II data with an exposure of 54 tonne-days, and
derived an upper limit on the mixing parameter $\lesssim10^{-10}$. The
XENON collaboration also carried out searches for a similar
model~\cite{Aprile:2019xxb}.
  
In this work, we perform searches for dark matter models with a light
mediator using the full dataset of PandaX-II, corresponding to a total
exposure of $132$ tonne-days, and find no significant excess over
background. We interpret the results in terms of a well-motivated SIDM
model that has been shown to explain dark matter distributions from
dwarf galaxies to galaxy clusters~\cite{Kaplinghat:2015aga,Huo:2017vef}. In addition, we
combine the PandaX-II constraints and those from observations of the light element abundances in the early
universe~\cite{Hufnagel:2018bjp,Depta:2020zbh}, and show that they
provide a complementary test of the self-interacting nature of dark
matter. In particular, our results indicate that the dark sector needs
to be colder than the visible sector in the early universe for dark matter masses ranging from
$\sim10~{\rm GeV}$ to $200~{\rm GeV}$.

\section{Experiment and Method}
\label{sec:2}

PandaX-II is a dark matter direct detection experiment, located in
China Jinping underground laboratory. Its central apparatus is a
dual-phase xenon time projection chamber, containing 580 kg liquid
xenon in the sensitive volume. Energy deposit from particles
interacting with liquid xenon results in prompt scintillation lights
($S1$ signal) and ionization electrons which are drifted up to gaseous
xenon and produce delayed second scintillation lights ($S2$
signal). Both signals are recorded in one trigger window (one
event). More detailed descriptions of the experiment can be found
in~\cite{Tan:2016diz,Tan:2016zwf,Cui:2017nnn}.

As in~\cite{Ren:2018gyx}, we first consider a general case where the
interaction between dark matter and nucleon is mediated by a force
carrier, $\phi$. If we further assume equal effective couplings to
proton and neutron as in the standard WIMP search, the general form of
the dark matter-nucleus elastic scattering cross section can be
parametrized as~\cite{DelNobile:2015uua}
\begin{equation}
  \sigma(q^2)_{\chi N} =
  \sigma|_{q^2=0}A^{2}\left(\frac{\mu}{\mu_{p}}\right)^{2}\frac{m^{4}_{\phi}}
         {(m^{2}_{\phi}+q^{2})^2}F^{2}(q^2),
         \label{eq:1}
\end{equation}
where $\sigma|_{q^2=0}$ is the dark matter-nucleon cross section in
the limit of zero momentum transfer ($q^2=0$), $A$ is the mass number
of the nucleus, $\mu$ ($\mu_{p}$) is the dark matter-nucleus (nucleon)
reduced mass, $m_\phi$ is the mediator mass, and $F(q^2)$ is the
nuclear form factor~\cite{Lewin:1995rx}. We can see that $\sigma_{\chi
  N}$ is momentum-dependent and it approaches the standard WIMP case
when $m_\phi\gg q$. The corresponding differential recoil rate
is~\cite{Savage:2008er}
  \begin{equation}
    \frac{dR}{dE} = \frac{\sigma(q^2)_{\chi N}\rho}{2m_{\chi}\mu^{2}}\int_
         {v\geq  v_{\mathrm{min}}}d^{3}v v f(v,t),
    \label{eq:2}
  \end{equation}
where $\rho=0.3~{\rm GeV/cm^{3}}$ is the local dark matter density,
$m_{\chi}$ is the dark matter mass, $f(v,t)$ is the time-dependent
velocity distribution relative to the detector, and $v_{\mathrm{min}}$
is the minimum velocity that results in a recoil energy of $E$. The
event rate is calculated based on the standard isothermal halo
model~\cite{Smith:2006ym,Savage:2008er} with a circular velocity of
$220~{\rm km/s}$, a galactic escape velocity of $544~{\rm km/s}$, and
an average Earth velocity of $245~{\rm km/s}$.
 
\begin{figure}[H]
    \centering
\includegraphics[scale=0.45]{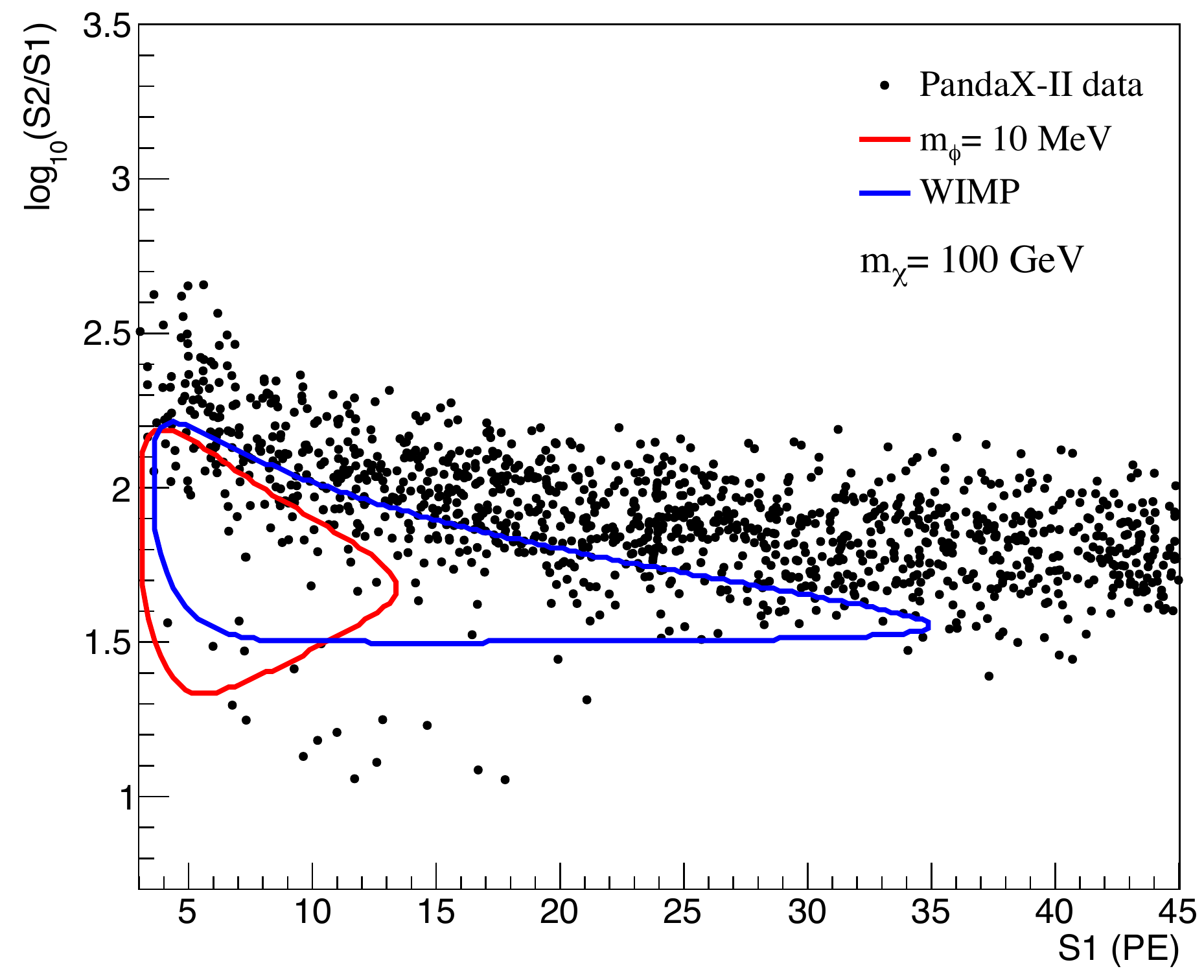}
    \caption{Selected dark matter candidate events in the $S1-\mathrm{log}_{10}(S2/S1)$ plane from the PandaX-II full dataset. The contours contain $68.3\%$ of the expected signals for a dark matter model with a light mediator where $m_\chi=100~{\rm GeV}$ and $m_{\phi}=10~{\rm MeV}$ (red) and a WIMP model with the same $m_\chi$ (blue).}
    \label{fig:datadistr}
  \end{figure}

We select dark matter candidate events in the same way as in the final
WIMP search of PandaX-II~\cite{Wang:2020coa}. The main event selection
requires one $S1$ signal in the range of $3$ photoelectron (PE) and
$45$ PE, and one $S2$ signal in the range of $100$ PE (uncorrected)
and $10000$ PE. ~\cref{fig:datadistr} shows the distribution of the
selected events in the $S1-\mathrm{log}_{10}(S2/S1)$ plane. The majority of events
are distributed in the band of $1.5<\mathrm{log}_{10}(S2/S1)<2.5$, which are dominated by 
electron recoil backgrounds. Some are in the region $\mathrm{log}_{10}(S2/S1)<1.5$ and $S1<20$, and they
are mostly from surface events; see~\cite{Wang:2020coa} for details. No significant excess events
in data was observed above the background prediction. Expected dark
matter yields are estimated using the above recoil event rate and the
updated NEST simulation framework~\cite{Szydagis:2018}. Parameters
used in the simulation, such as the light and charge yield, are tuned
against the PandaX-II calibration data. Measured experimental
efficiencies are also applied. \cref{fig:datadistr} also shows that
the light-mediator model and the traditional WIMP model have different
signal regions due to the $m_{\phi}$ dependence of the recoil energy
spectrum.

For statistical analysis, we follow the same construction of test
statistics as in Ref.~\cite{Wang:2020coa} which is based on profile
likelihood ratio. Then the standard $CL_{s+b}$
approach~\cite{Junk:1999kv} is used to derive the upper limits of
$\sigma|_{q^2=0}$ given dark matter and mediator masses. In the case
of strong downward fluctuation, the reported limits are
power-constrained to $-1\sigma$ of the sensitivity band evaluated from
background-only toy data sets.
  
  \begin{figure*}[!htp]\centering
    \includegraphics[width=0.45\linewidth]{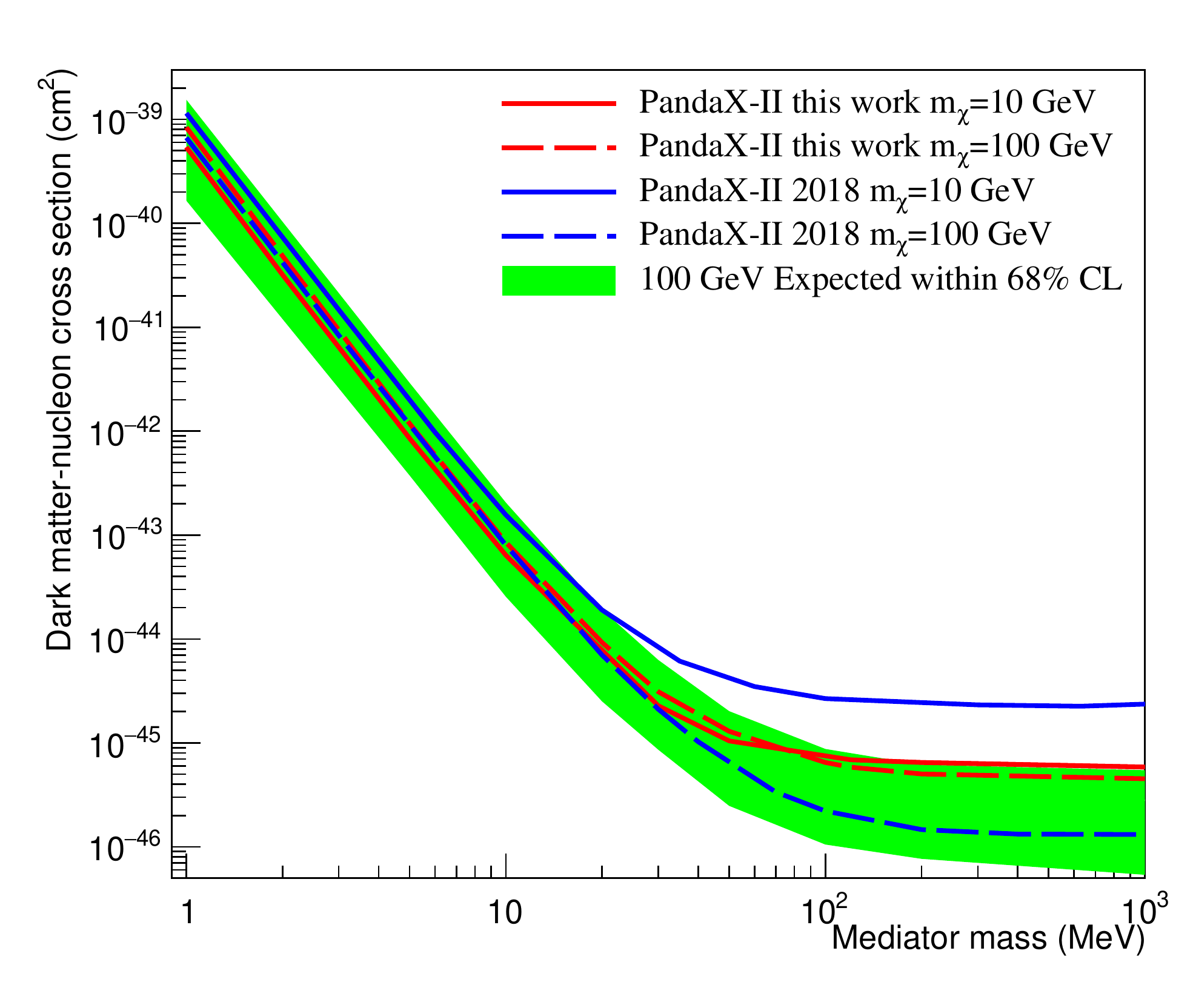}\hspace{1cm}
    \includegraphics[width=0.45\linewidth]{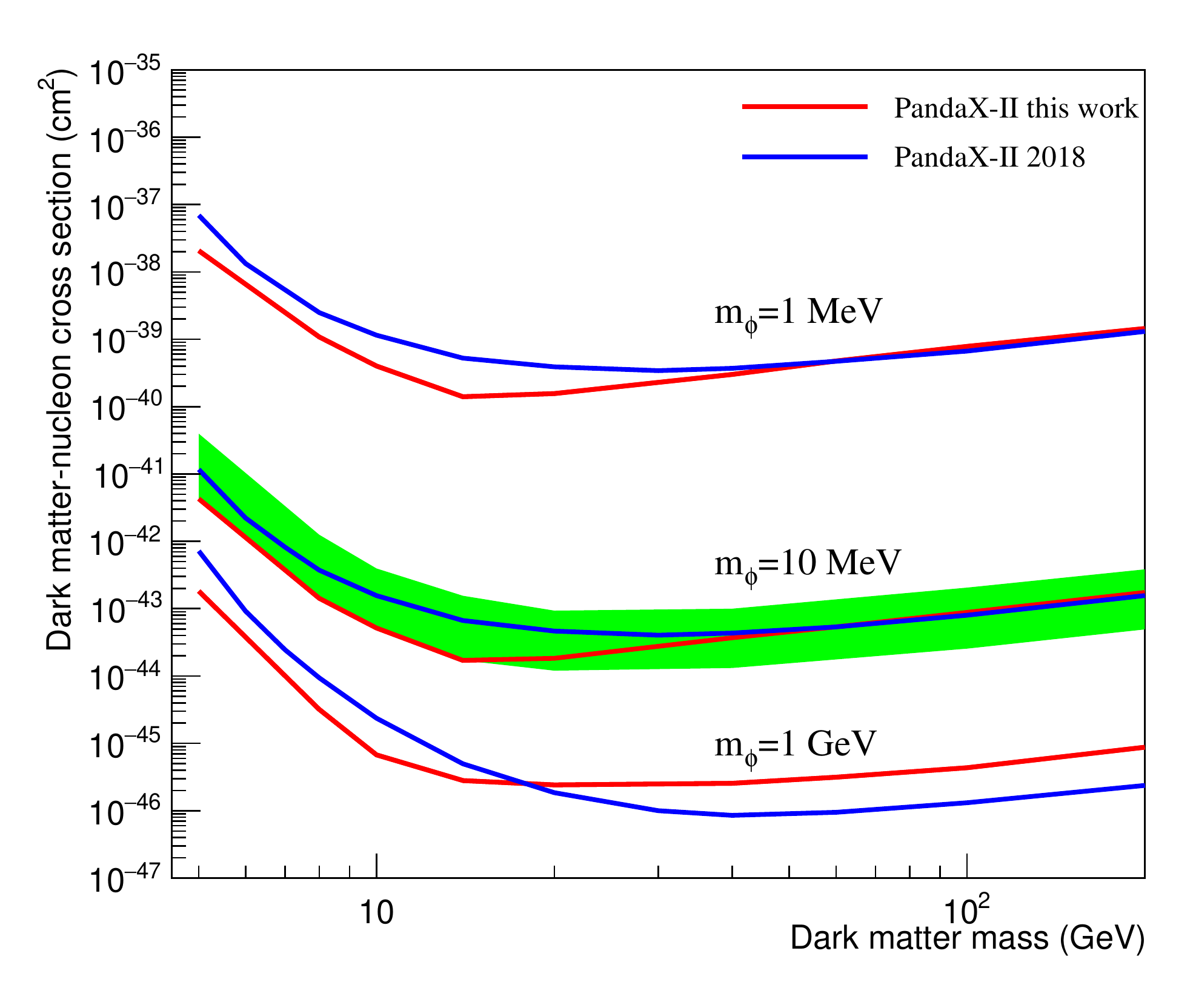}
    \caption{{\it Left:} PandaX-II 90\% CL upper limits on the zero-momentum dark matter-nucleon cross section vs the mediator mass for $m_\chi=10~{\rm GeV}$ (red solid) and $100~{\rm GeV}$ (red dashed). {\it Right:} the limits vs the dark mass for$m_\phi=1~{\rm MeV}$, $10~{\rm MeV}$ and $1~{\rm GeV}$. In both panels, the green band denotes the $\pm1\sigma$ sensitivity for the given model parameters. For comparison, the limits from our previous analysis based on the PandaX-II data release in 2018~\cite{Ren:2018gyx} are shown (blue).}
    \label{fig:lim_xsmphi}
  \end{figure*}

\section{Results and Discussion}
\label{sec:3}
  
The left panel of \cref{fig:lim_xsmphi} shows the 90\% CL upper limits
on the $\sigma|_{q^2=0}$ vs the mediator mass for $m_\chi=10~{\rm
  GeV}$ (red) and $100~{\rm GeV}$ (blue). Overall, they become
stronger as the mediator mass increases. For $m_\phi\gtrsim100~{\rm
  MeV}$, our limits approach those derived for the WIMP
case~\cite{Wang:2020coa}, since the momentum transfer becomes
negligible in Eq.~\ref{eq:1}. For $m_\chi=10~{\rm GeV}$, our current
exclusion limits are stronger than those found in the 2018 analysis
for all $m_\phi$. For $m_\chi=100~{\rm GeV}$, both limits are comparable 
for $m_{\phi}\lesssim30~{\rm MeV}$. The right panel of \cref{fig:lim_xsmphi} shows the limits vs the dark
matter mass. In the WIMP case where $m_\phi=1~{\rm GeV}$, the current
limits are stronger for $m_\chi\lesssim20~{\rm GeV}$. For
$m_\phi=1~{\rm MeV}$ and $10~{\rm MeV}$, our current limits are
stronger for $m_\chi\lesssim40~{\rm GeV}$. Consider $m_\chi\approx15~{\rm GeV}$
and $m_\phi=10~{\rm MeV}$, we find $\sigma|_{q^2=0}<1.7\times
10^{-44}~{\rm cm^{2}}$, which is a factor of $3.9$ improvement
compared to the 2018 analysis.
  
To further interpret our search results, we consider a well-motivated
SIDM model where dark matter is assumed to be a Dirac fermion and it
couples to a vector mediator with a gauge
coupling. Refs.~\cite{Kaplinghat:2015aga,Huo:2017vef} show that this model can explain dark matter distributions inferred from  spiral galaxies and galaxy clusters, and both mediator
mass and gauge coupling can be determined for a range of dark matter masses.
Here, we further assume the mediator couples to the photon
through kinetic mixing~\cite{Holdom:1985ag},
$\epsilon_{\gamma}\phi_{\mu\nu}F^{\mu\nu}$, where $\epsilon_{\gamma}$
is the mixing parameter, $\phi_{\mu\nu}$ and $F^{\mu\nu}$ are the
mediator and photon field strengths, respectively. In the $q^{2}=0$
limit, the dark matter-nucleon cross section is
  \begin{equation}
    \sigma|_{q^2=0} = \frac{16\pi\alpha_{\rm EM}\alpha_{\chi}\mu^{2}_{p}}
           {m^{4}_{\phi}}\left[\frac{\epsilon_{\gamma}Z}{A}\right]^{2},
           \label{eq:sigma0}  
  \end{equation}
where $\alpha_{\rm EM}=1/137$ and $\alpha_\chi$ are the fine
structure constants in the visible and dark sectors, respectively,
$Z$ is the proton number of the xenon nuclei. In our analysis, we fix
$\alpha_{\chi}$ to be the best-fit value for given $m_\chi$ as
in~\cite{Huo:2017vef}.
  
\begin{figure*}[!htbp]
   \centering \includegraphics[scale=0.4]{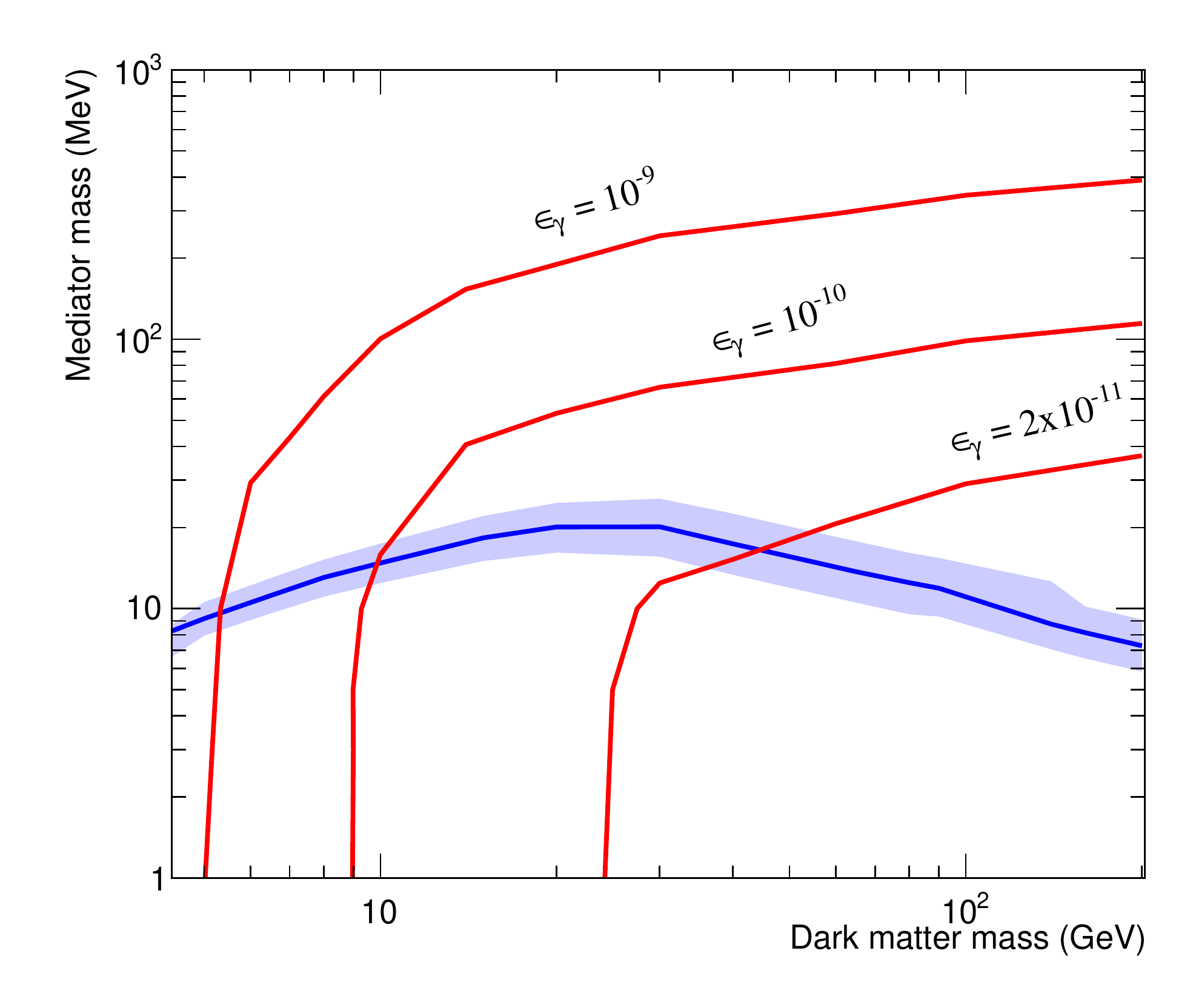}\hspace{1cm}
   \includegraphics[scale=0.4]{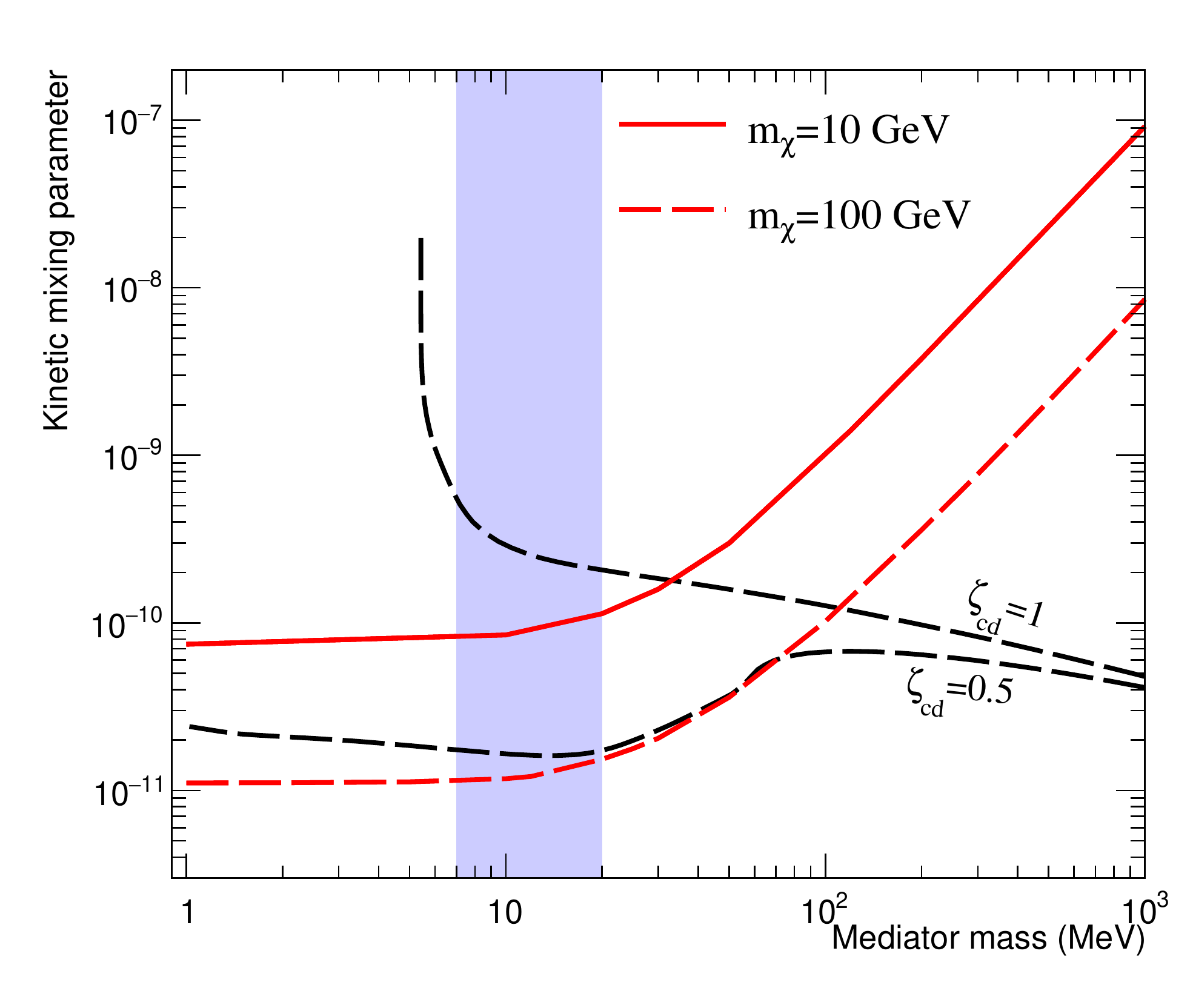}
  \caption{{\it Left:} PandaX-II lower limits on the mediator mass for
    the kinetic mixing parameter $\epsilon_{\gamma}=2\times10^{-11}$,
    $10^{-10}$ and $10^{-9}$ (red). The blue band denotes the SIDM
    parameter region favored by observations of dark matter halos from
    dwarf galaxies to galaxy clusters~\cite{Huo:2017vef}. {\it Right:}
    PandaX-II upper limits on the kinetic mixing parameter for
    $m_\chi=10~{\rm GeV}$ and $100~{\rm GeV}$ (red). The vertical band
    indicates the mediator mass range for SIDM. The black curves
    indicate lower limits on $\epsilon_\gamma$ from observations of
    the light element abundances in the early universe, assuming two
    benchmark values of the temperature ratio $\zeta_{\rm cd}=1$ and
    $0.5$; calculated based on the results in~\cite{Depta:2020zbh}.}
      \label{fig:mphilimit_fixed_alphax}
  \end{figure*}

The left panel of \cref{fig:mphilimit_fixed_alphax} shows the PandaX-II constraints in
the $m_{\phi}\textup{--}m_{\chi}$ plane for three representative
values of $\epsilon_\gamma$ (red), together with the favored SIDM
parameter region from astrophysical observations (blue), taken
from~\cite{Huo:2017vef}. Our results can constrain a large portion of
the SIDM parameter space even for $\epsilon_{\gamma}$ being as small
as $2\times10^{-11}$, at which the SIDM model with
$m_\chi\gtrsim40~{\rm GeV}$ is excluded. The sensitivity further
improves as $\epsilon_\gamma$ increases. For
$\epsilon_\gamma=10^{-9}$, the region with $m_\chi\gtrsim6~{\rm GeV}$
is excluded. Alternatively, we can derive 
upper limits in the $\epsilon_\gamma\textup{--}m_\phi$ plane for given $m_\chi$. This is shown in the right panel of ~\cref{fig:mphilimit_fixed_alphax} for two representative dark matter masses, $10~{\rm GeV}$ and $100~{\rm GeV}$ (red).

There are cosmological constraints on the model parameters. In the early universe, the mediator must
decay to (almost) massless particles to avoid over-closing the
universe~\cite{Kaplinghat:2013yxa,DelNobile:2015uua}. If it decays to
standard model particles via the kinetic mixing portal discussed
above, the energy injection to the plasma in the early universe may
alter the light element abundances and lead to tensions with
successful predictions of the standard Big Bang Nucleosynthesis (BBN)
scenario. Refs.~\cite{Hufnagel:2018bjp,Depta:2020zbh} study BBN constraints and
derive lower limits on the mediator lifetime $\tau_\phi$, which depend
on $m_\phi$, the chemical decoupling temperature of $\phi$ ($T_{\rm
  cd}$) and the temperature ratio when the decoupling occurs
($\zeta_{\rm cd}$). The last two parameters essentially control the
abundance of mediator particles in the early universe. We take their
limits for $T_{\rm cd}=10~{\rm GeV}$, $\zeta_{\rm cd}=1$ and $0.5$
from~\cite{Depta:2020zbh}, two benchmark examples, and convert them
into lower limits on $\epsilon_\gamma$ for a given $m_\phi$ via the
relation $\tau_{\phi}=3/(\alpha_{\mathrm{SM}}m_{\phi}\epsilon^{2}_\gamma)$~\cite{Kaplinghat:2013yxa},
which are shown in the right panel
of~\cref{fig:mphilimit_fixed_alphax} (black). Since higher $\zeta_{\rm cd}$
indicates higher $\phi$ abundances, its lifetime should be shorter to
avoid the BBN constraints, resulting in larger $\epsilon_\gamma$.

The joint PandaX-II and BBN constraints put interesting tests on the
SIDM model. If both sectors have the same temperature as the mediator decouples, i.e., $\zeta_{\rm cd}=1$, $m_\phi$ has to be higher than $35~{\rm MeV}$ and $110~{\rm MeV}$ for $m_\chi=10~{\rm GeV}$ and
$100~{\rm GeV}$, respectively. Thus for the mass range of dark matter we consider, the joint constraints strongly
disfavor the SIDM parameter region, as the required mediator mass is $m_\phi\sim10~{\rm MeV}$, denoted by the vertical band of the right panel of \cref{fig:mphilimit_fixed_alphax}. On the other hand, for $\zeta_{\rm
  cd}=0.5$, the SIDM region is allowed unless $m_\chi
\gtrsim100~{\rm GeV}$. Overall, our results show that the dark sector needs to
be colder than the visible sector in the early universe.

\section{Summary}
\label{sec:4}
  
We have used the PandaX-II full dataset to constrain dark matter
models with a light mediator and derived upper limits on the dark
matter-nucleon elastic scattering cross section. Compared to our
previous study based on the PandaX-II data release in 2018, the
improvement is significant for dark matter mass below $20~{\rm GeV}$. We
interpreted our results in terms of an SIDM model, which is motivated
to explain dark matter distributions from dwarf galaxies to galaxy
clusters, and obtained stringent upper limits on the kinetic mixing
parameter. We further derived joint constraints on the SIDM model by
taking into account {\em both} PandaX-II searches and observations of
the light element abundances in the early universe. Our results
demonstrate that direct detection, astrophysical and cosmological
observations provide complementary searches for dark matter
self-interactions. The next generation of the PandaX experiment,
PandaX-4T~\cite{Zhang:2018xdp}, is expected to improve the detection
sensitivity by more than one order of magnitude, and it will further
test the particle nature of dark matter.

  \Acknowledgements{ This project has been supported by a Double
    Top-class grant from Shanghai Jiao Tong University, grants from
    National Science Foundation of China (No 11875190). We thank the
    Office of Science and Technology, Shanghai Municipal Government
    and the Key Laboratory for Particle Physics, Astrophysics and
    Cosmology, Ministry of Education, for important support. This work
    is supported in part by the Chinese Academy of Sciences Center for
    Excellence in Particle Physics (CCEPP) and Hongwen Foundation in
    Hong Kong. Finally, we thank the CJPL administration and the
    Yalong River Hydropower Development Company Ltd. for indispensable
    logistical support and other help. HBY acknowledges support from
    the U. S. Department of Energy under Grant No. de-sc0008541 and
    the John Templeton Foundation under Grant ID \# 61884.}

\end{multicols}
\end{document}